\documentclass[A4]{pasj00}

\Received{$\langle$reception date$\rangle$}
\Accepted{$\langle$acception date$\rangle$}
\Published{$\langle$publication date$\rangle$} \SetRunningHead{Erlin
Qiao \&  B.F. Liu}{The Spectral Features of Diak and Corona with Mass Evaporation in the Low/Hard State}
\usepackage{times}
\usepackage{graphicx}
\usepackage{epsfig}
\usepackage{lscape}

\begin{document}

\title{The Spectral Features of Disk and Corona with Mass Evaporation in the Low/Hard State}
\author{Erlin Qiao\altaffilmark{1,2,3} and B.F. Liu\altaffilmark{1,3,}}
\affil{%
\altaffilmark{1} National Astronomical Observatories /Yunnan Observatory,
Chinese Academy of Sciences, P.O. Box 110, Kunming 650011, P. R. China\\
\altaffilmark{2} Graduate School of Chinese Academy of Sciences, Beijing 100049, P. R. China\\
\altaffilmark{3} Key Laboratory for the Structure and Evolution of Celestial Objects, Chinese Academy of Sciences}

\email{qel1982@ynao.ac.cn \\
bfliu@ynao.ac.cn}

\maketitle

\begin{abstract}
We investigate the spectral features of accretion flows composed of an outer cool, optically thick disk and inner
hot, optically thin, advection dominated accretion flows (ADAF) within the
framework of disk and corona with mass evaporation (Liu et al. 2002a). In this work, both the magnetic field
and Compton scattering of soft photons from the disk by electrons in the corona are included to calculate the
 evaporation rates at different distances. The disk is truncated at the distance where the evaporation rate equals to the accretion
rate ($\dot m_{\rm evap}(r_{\rm tr})=\dot m$). For a series of accretion rates, the corresponding truncation
radii are calculated out, with which we are able to calculate the emergent spectra from the inner ADAF + outer disk + corona.
At very low accretion rates, the spectra are similar to that of a pure ADAF because the disk is truncated at
large distances. The disk component becomes important at high accretion rates since the truncation occurs at small distances.
An anti-correlation between the Eddington ratio $\xi \equiv L_{\rm 0.5-25 \,keV}/L_{\rm Edd}$ and the
hard X-ray photon index $\Gamma_{\rm 3-25 \,keV}$ at low/hard states is predicted by the model.
Comparing the theoretical results with observations, we find that our model can reproduce the
anti-correlation between the Eddington ratio $\xi$ and the hard X-ray photon index  observed for
the X-ray binary XTE J1118+480.
\end{abstract}

\section{Introduction}
Accretion onto stellar black hole is the most widely accepted
scenario for reproducing the different spectra in black hole X-ray
binaries. Generally, there are two basic spectral states, i.e. the
high/soft state and the low/hard state (Esin et al. 1997; Van de
Klis 1994; Nowak 1995; Tanaka \& Lewin 1995; Tanaka \& Shibazaki
1996; for reviews see Remillard \& McClintock 2006 and Done et al.
2007). In the high/soft state, the accretion occurs dominantly via a
cool, optically thick disk (Pringle \& Rees 1972; Shakura \& Sunyaev
1973; Mitsuda et al. 1984; Frank et el. 2002), while in the low/hard
state, the accretion dominantly via a hot, optically thin advection
dominated accretion flow (ADAF) or radiative inefficient accretion
flow (RIAF) (Rees et al. 1982; Narayan \& Yi 1994, 1995a, b;
Abramowicz et al. 1995; Narayan 2005; Kato et al. 2008). Detailed
fits to the observational spectra show that in the high/soft state
the disk extends to the innermost stable circular orbit (ISCO),
while in the low/hard state, the disk truncates at some radius, and
from the truncation radius inward the disk is replaced by an ADAF or
RIAF (Kawabata \& Mineshige 2010). The spectra in the low/hard state
is characterized by a hard power law ($\Gamma \sim 1.5-2.1$) peaking
at $\sim$ 100 keV, and sometimes accompanied by a low-temperature
thermal component (Esin et al. 2001; McClintock et al. 2001; Done et
al. 2007). The hard power-law component is believed to be produced
by the inner hot ADAF or RIAF, and the thermal component by the
truncated disk (Esin et al. 2001; Done et al. 2007).

Several theoretical models are proposed to be potential mechanisms for
the disk truncation in the low/hard state for black hole
X-ray binaries (Honma 1996; Meyer, Liu \& Meyer-Hofmeister 2000b; Manmoto \& Kato 2000; Lu et al. 2004;
R\`o\.za\`nska \& Czerny 2000; Spruit \& Deufel 2002; Dullemond \& Spruit 2005).
The disk-corona evaporation model (Meyer \& Meyer-Hofmeister 1994;
Meyer et al. 2000a, b; Liu et al. 2002a)
can naturally explain the transition between the outer thin disk and the inner ADAF/RIAF.
The interaction between the disk and the corona leads to mass
evaporating from the disk to the corona.
When the mass accretion rate in the disk is lower than the evaporation rate from the disk to the corona, the disk
is truncated. For given mass of black hole and accretion rate, the truncation radius is determined by the
evaporation model. Therefore, we can calculated  the radiative spectra from the inner ADAF, the outer disk and corona.

Thanks to the rapid progress of the observational techniques, more
detailed data for the multi-wavelength spectra are available for
studying the evolution of the truncation radius of the disk with
mass accretion rate in black hole X-ray binaries (Cabanac et al.
2009). The evolution of X-ray spectra with mass accretion rate has
been also investigated in both AGNs and black hole X-ray binaries
(Esin et al. 1997; wang et al. 2004; Gu et al. 2009; Wu et al. 2008;
Liu et al. 2009). Here, for the first time, we calculate the
emergent spectra from the disk and corona, and the inner ADAF caused
by the mass evaporation.  We then compare the theoretical spectra of
our model for different mass accretion rates with observations.

In this work, we perform more detailed calculations to study the evaporation and disk truncation in the low/hard state.
Both the cooling by Compton scattering of the disk photons
and the magnetic field are included to calculate the evaporation rates, which makes our solution more
self-consistent. We obtain a fitting formula for the truncation radius as a function of mass accretion rates. Then,
we calculate the emergent spectra from the inner ADAF, the outer disk and corona for different mass
accretion rates. The dependence of the hard X-ray photon index on the mass accretion rate is derived.
From the theoretical spectra, we find an
anti-correlation between the Eddington ratio $\xi \equiv L_{\rm 0.5-25 \,keV}/L_{\rm Edd}$ and
the hard X-ray photon index $\Gamma_{\rm 3-25 \,keV}$, which is consistent with observations
for the black hole X-ray binary XTE J1118+480 (Wu et al. 2008).

The paper is organized as follows. In section 2 we briefly describe the disk-corona model. In section 3 we present the numerical
results, and in section 4 comparison with observations. The
discussions and conclusions are given in section 5 and section 6.

\section{The Model}
We consider a hot corona above a geometrically thin standard disk
around a central black hole. In the corona, viscous dissipation
leads to ion heating, which is partially transferred to the
electrons by means of Coulomb collisions. This energy is then
conducted down into lower, cooler and denser corona. If the density
in this layer is sufficiently high, the conductive flux is radiated
away. If the density is too low to efficiently radiate the energy,
cool matter is heated up and evaporation into the corona takes
place. The mass evaporation goes on until an equilibrium density is
established. The gas evaporating into the corona still retains
angular momentum and with the role of viscosity will differentially rotate
around the central object. By friction the gas looses angular momentum and drifts
inward thus continuously drains mass from the corona towards the
central object. This is compensated by a steady mass evaporation
flow from the underlying disk. The process is driven by the
gravitational potential energy released by friction in the form of
heat in the corona. Therefore, mass is accreted to the central
object partially through the corona (evaporated part) and partially
through the disk (the left part of the supplying mass). Such a model
was proposed by Meyer \& Meyer-Hofmeister (1994) for dwarf novae,
established for black holes by Meyer et al. (2000a) and modified by
Liu et al. (2002a) where the decoupling of ions and electrons and
Compton cooling effect are taken into account. Further updates by
Meyer-Hofmeister \& Meyer (2003), Qian et al. (2007) and Qiao \& Liu (2009) are also
included in this study. For clarity, we list the equations
describing the physics of corona as follows.

Equation of state
\begin{equation}
\centering
 P={\Re \rho \over 2\mu}(T_{\rm i}+T_{\rm e})\beta^{-1},
\end{equation}
where $\mu=0.62$ is the molecular weight assuming a standard
chemical composition ($X=0.75, Y=0.25$) for the corona, $\beta$ is
the ratio of gas pressure $P_{\rm g}$ to total pressure $P$
$(P=P_{\rm g}+P_{\rm m}$, where $P_{\rm m}$ is magnetic pressure).
For convenience, we assume the number density of ion $n_{\rm i}$
equals to that of electron $n_{\rm e}$, which is strictly true only
for a pure hydrogen plasma.

Equation of continuity
\begin{equation}\label{e:continue}
\centering
 {d\over dz}(\rho v_{\rm z})=\eta_{\rm M}{2\over R}\rho v_{\rm R} -{2z\over
R^2+z^2}\rho v_{\rm z},
\end{equation}

where $v_{\rm R}=-\alpha {V_{\rm s}^2 \over v_{\rm \varphi}}$ is the radial component of velocity,
$V_{\rm s}=({P \over \rho})^{1/2}$ is the isothermal sound speed,
$v_{\rm \varphi}=\sqrt{G M \over r} (1+{z^2 \over r^2})^{-4/3}$
is the angular component of velocity (for the detailed derivation of $v_{\rm \varphi}$,
one can see Meyer et al. 2000b).

Equation of the $z$-component of momentum
\begin{equation}\label{e:mdot}
\rho v_{\rm z} {dv_{\rm z}\over dz}=-{dP\over dz}-\rho {GMz\over
(R^2+z^2)^{3/2}},
\end{equation}

The energy equation of ions
\begin{equation}\label{e:ions}
\begin{array}{l}
{d\over dz}\left\{\rho_{\rm i} v_{\rm z} \left[{v^2\over 2}+{\gamma\over
\gamma-1}{P_{\rm i}\over \rho_{\rm i}}-{GM\over (R^2+z^2)^{1\over
2}}\right]\right\}\\
={3\over 2}\alpha P\Omega-q_{\rm ie}\\
+{\eta_{\rm E}}{2\over R}\rho_{\rm i} v_{\rm R}
\left[{v^2\over 2}+{\gamma\over \gamma-1}{P_{\rm i}\over \rho_{\rm i}}-{GM\over (R^2+z^2)^{1\over 2}}\right]\\
-{2z\over {R^2+z^2}}\left\{\rho_{\rm i} v_{\rm z} \left[{v^2\over
2}+{\gamma\over \gamma-1}{P_{\rm i}\over \rho_{\rm i}}-{GM\over
(R^2+z^2)^{1\over 2}}\right]\right\},
\end{array}
\end{equation}

where $\eta_{\rm M}$ is the mass advection modification term and $\eta_{\rm E}$
is the energy modification term. We take $\eta_{\rm M}=1$ for the case
without consideration of the effect of mass inflow and outflow from
and into neighboring zones in the corona, and $\eta_{\rm E}=\eta_{\rm M}+0.5$ is
a modification to previous energy equations (for details see
Meyer-Hofmeister \& Meyer 2003). $\Omega$ is the angular velocity
of the corona, $\Omega= \sqrt{G M \over r^3} (1+{z^2 \over r^2})^{-4/3}$
(for the detailed derivation of $\Omega$, one can see Meyer et al. 2000b).
$q_{\rm ie}$ is the exchange rate of
energy between electrons and ions,

\begin{equation}
{q_{\rm ie}}={\bigg({2\over \pi}\bigg)}^{1\over 2}{3\over 2}{m_{\rm e}\over
m_{\rm i}}{\ln\Lambda}{\sigma_{\rm T} c n_{\rm e} n_{\rm i}}(\kappa T_{\rm i}-\kappa T_{\rm e})
{{1+{T_{\rm *}}^{1\over 2}}\over {{T_{\rm *}}^{3\over 2}}},
\end{equation}

with
\begin{equation}
T_{\rm *}={{\kappa T_{\rm e}}\over{m_{\rm e} c^2}}\bigg(1+{m_{\rm e}\over m_{\rm i}}{T_{\rm i}\over
T_{\rm e}}\bigg),
\end{equation}
where $m_{\rm i}$ and $m_{\rm e}$ are the proton and electron masses, $\kappa$ is the
Boltzmann constant, $c$ is the light speed, $\sigma_{\rm T}$ is the Thomson
scattering cross section and $\ln\Lambda=20$ is the Coulomb logarithm.

The energy equation for both the ions and electrons
\begin{equation}\label{e:energy}
\begin{array}{l}
{\frac{d}{dz}\left\{\rho {v}_{\rm z}\left[{v^2\over
2}+{\gamma\over\gamma-1}{P\over\rho}
-{GM\over\left(R^2+z^2\right)^{1/2}}\right]
 + F_{\rm c} \right\}}\\
=\frac{3}{2}\alpha P{\mit\Omega}-n_{\rm e}n_{\rm i}L(T)-q_{\rm Comp}\\
+\eta_{\rm E}{2\over R}\rho v_{\rm R} \left[{v^2\over
2}+{\gamma\over\gamma-1}{P\over\rho}
-{GM\over\left(R^2+z^2\right]^{1/2}}\right]\\
-{2z\over R^2+z^2}\left\{\rho v_{\rm z}\left[{v^2\over
2}+{\gamma\over\gamma-1}{P\over\rho}-
{GM\over\left(R^2+z^2\right)^{1/2}}\right] +F_{\rm c}\right\},
\end{array}
\end{equation}

where $n_{\rm e}n_{\rm i}L(T)$ is the bremsstrahlung cooling rate and $F_{\rm c}$
is the thermal conduction (Spitzer 1962),
\begin{equation}\label{e:fc}
F_{\rm c}=-\kappa_{\rm 0}T_e^{5/2}{dT_{\rm e}\over dz},
\end{equation}
with $\kappa_0 = 10^{-6}{\rm erg\,s^{-1}cm^{-1}K^{-7/2}}$ for fully
ionized plasma.

$q_{\rm Comp}$ is the Compton cooling rate,
\begin{equation}
q_{\rm Comp} = {\frac{4\kappa T_{\rm e}}{m_{\rm e} c^2}}n_{\rm e} \sigma_{\rm T} c {\frac{a
T_{\rm eff} ^4}{2}},
\end{equation}
with $T_{\rm eff}$ the effective temperature of the underlying thin
disk and $a$ the radiation constant.
\begin{equation}
T_{\rm eff}=\frac{3GM\dot M_{\rm d}}{8\pi R^3 \sigma},
\end{equation}
where  $\dot M_{\rm d} (R)$ is the mass accretion rate in the thin
disk, which depends on the distance because of evaporation,
\begin{equation}
\dot M_{\rm d} (R)= \dot M_{} -\dot M_{\rm evap} (R),
\end{equation}
With $\dot M_{}$ the mass transfer rate from the secondary, which is assumed equal to the mass accretion
rate in the most outer region of the accretion disk
and $\dot M_{\rm evap}$ the integrated evaporation rate
from the outer edge of the disk to the distance $R$.

All other parameters in above equations are under standard
definitions, and are in cgs units. The five differential equations,
Eqs. (\ref{e:continue}), (\ref{e:mdot}), (\ref{e:ions}),
(\ref{e:energy}), (\ref{e:fc}), which contain five variables $P(z)$,
$T_{\rm i}(z)$, $T_{\rm e}(z)$, $F_{\rm c}(z)$, and $\dot m(z)(\equiv
\rho v_{\rm z})$ can be solved with five boundary conditions.

At the lower boundary $z_0$ (the interface of disk and corona), the
temperature of the gas should be the effective temperature of the
accretion disk. Previous investigations (Liu, Meyer, \&
Meyer-Hofmeister 1995) show that the coronal temperature increases
from effective temperature to $10^{6.5}$K in a very thin layer and
thus the lower boundary conditions can be reasonably approximated
(Meyer et al. 2000a; Qian et al. 2007) as,
\begin{equation}
\begin{array}{l}
{T_{\rm i}=T_{\rm e}=10^{6.5}} \,{\rm K},\\
F_{\rm c}=-2.73\times 10^6 \beta P \ {\rm at}\ z=z_0.
\end{array}
\end{equation}
At infinity, there is  no pressure and no heat flux. This requires
sound transition at some height $z=z_1$. We then constrain the upper
boundary as,
\begin{equation}
F_{\rm c}=0\  {\rm and}
 \  v_{\rm z}=V_{\rm s}\ {\rm at}\ z=z_1.
\end{equation}
With such boundary conditions, we assume a set of lower boundary
values for $P$ and $\dot m$ to start the integration along $z$. Only
when the trial values for $P$ and $\dot m$ fulfill the upper
boundary conditions can the presumed $P$ and $\dot m$ be taken as
true solutions of the differential equations.

\section{Numerical Results}
In our calculation we fix the central black hole mass as
$M=6M_\odot$,  viscosity parameter as $\alpha=0.3$, and assume
equipartition of gas pressure and magnetic pressure, i.e.,
$\beta=0.5$. The evaporation rate is
calculated from the vertical mass flow at the lower boundary,
$\dot m_0\equiv \rho v_{\rm z}$,
and is then integrated in the radial one-zone region, $\dot M_{\rm
evap}\approx 2\pi R^2\dot m_0$. This integrated evaporation rate represents
the mass flowing rate in the local corona if there is no gas supply
except for the mass transfer from the secondary in a binary system.
If the mass transfer rate $\dot M_{}$ is less than the maximum evaporation rate,
the thin disk will truncate at the radius $R_{\rm tr}$ where the
accretion rate equals to the evaporation rate. The region from the truncation
radius inward is filled with a hot ADAF and the region from the truncation radius outward
there is a thin disk, above which there is a hot corona
providing the material for the ADAF. If the mass transfer rate $\dot M_{}$ is
higher than the maximal evaporation rate, the thin disk will extend down
to the ISCO, only a very weak corona exists above the thin disk.

\subsection{The Evaporation Features and Disk Truncation}
Figure 1 shows the evaporation curves for different mass accretion
rates. Here the evaporation rate is scaled with the Eddington
accretion rate, $\dot M_{\rm Edd}=L_{\rm Edd}/{\eta c^2}= 1.39\times
10^{18} M/M_\odot\,{\rm g\,s^{-1}}$ (where $\eta=0.1$ is the energy
conversion efficiency) and the radius is scaled with the
Schwarzschild radius, $R_ {\rm S}=2GM/c^2=2.95 \times 10^5 M/M_\odot
\, {\rm cm}$. As the first step, we neglect the effect of Compton
cooling in the corona associated with soft photons emitted by the
underlying thin disk. The evaporation rate without Compton cooling
is shown by the solid curve in Figure 1. Taking into account the
Compton effect, we calculate the evaporation rates for a series of
accretion rates. For $\dot M = 0.005 \,\dot M_{\rm Edd}$, the
evaporation curve is almost the same as the case without considering
the Compton cooling, which can be seen from Figure 1 for the long-dashed line.
We get an analytical formula between the mass accretion rate and the
truncation radius by fitting the numerical results without Compton cooling,
which is suitable for the case of $\dot M < 0.005 \,\dot M_{\rm Edd}$, that is,
\begin{equation}\label{e:tr1}
r_{\rm tr}= 2.215 \dot m^{-0.776}
\end{equation}
where $\dot m$ is scaled with Eddington accretion
rate and $r_{\rm tr}$ is scaled with Schwarzschild radius. Using this formula,
the truncation radius is $200 \,R_{\rm S}$ for $\dot M=0.003 \,\dot M _{\rm Edd}$ and
$471 \,R_{\rm S}$ for $\dot M= 0.001 \,\dot M _{\rm Edd}$ respectively.

With increase of the mass accretion rate, the evaporation is
suppressed  by the strong Compton cooling. For $\dot M= 0.008 \,\dot M_{\rm Edd}$,
the evaporation curve deviates the case without considering the Compton cooling, and
the evaporation rate becomes weaker. However,
the deviation is slight and the thin disk truncates at around $78.5 \,R_{\rm S}$, which is
not quite different from the case without the Compton cooling, one can see the dot-dashed line
from Figure 1. Increasing the mass accretion rate further to $\dot M= 0.01 \,
\dot M _{\rm Edd}$, the evaporation becomes very weak, and the thin disk truncates at a relative small
radius $40 \,R_{\rm S}$. For clarity, we show the relation
between the mass accretion rate and the truncation radius with the Compton cooling of
the soft photos from the thin disk to the corona
in Figure 2. The truncated disk corona plus ADAF geometry described above
corresponds to the low/hard state.

In order to check whether the inner ADAF solution can connect with the outer
disk-corona solution smoothly, we take $\dot M = 0.01 \,\dot M_{\rm Edd}$
as an example to demonstrate the distribution of the temperature along radial direction.
The thin disk truncates at about $40 \,R_{\rm S}$ for $\dot M= 0.01 \,\dot M_{\rm Edd}$.
From Figure 3, we can see the ion
temperature of the outer corona connect with the inner ADAF
well (the dashed line). The radial distribution of ion temperature is similar to the virial
temperature ($T_{\rm i} \propto R^{-1}$) (the solid line). However, the electron
temperature of the outer corona is lower than that of the inner
ADAF (the dotted line). This can be understood as follows. Because of the
presence of the cool thin disk underneath, the cooling of the corona
by electron thermal conduction between the corona and the underlying
thin disk is much more efficient than that of the radiative cooling,
thereby leads to a lower electron temperature of the corona than that of the ADAF.

With increase of the mass accretion rate further for $\dot M= 0.011 \,
\dot M_{\rm Edd}$, the evaporation rate decreases severely.
Because the peak value of the evaporation rate is
less than $0.011 \,\dot M_{\rm Edd}$, the thin disk extends to the
ISCO, and only a weak corona exists above the disk.
we can see the dotted line from Figure 1.
This configuration corresponds to the high/soft state, and this
accretion rate corresponds to the critical accretion rate between the
low/hard state and high/soft state transition. We note that the
critical accretion rate $\dot M= 0.011 \,\dot M_{\rm Edd}$ is less
than the previous result, $\dot M \approx 0.03 \,\dot M_{\rm Edd}$,
from our model for $\alpha=0.3$ (Meyer et al. 2000a, b; Liu et al. 2002a; Qiao \& Liu 2009).
 This is because the Compton cooling of the soft photons from the thin disk to the corona is considered.
The inverse Compton scattering will make the cooling in the corona more efficient,
and then suppresses the evaporation, thereby leading to a relative
smaller evaporation rate.

\subsection{The Spectra of Disk and Corona in the Low State}
If $\dot M < 0.01 \,\dot M_{\rm Edd}$, the thin disk is truncated by
mass evaporation. With determination of truncation radius as
described in section 3.1, we can calculate the emergent spectra from
the inner ADAF/RIAF and outer disk and corona. We describe the
ADAF/RIAF by the self-similar solution (Narayan et al. 1994, 1995b),
and the region from the truncation radius outward by two phase
disk-corona accretion flows. The physical quantities for calculating
the emergent spectra are the distribution of the density $\rho$,
Thomson scattering optical depth $\tau_{\rm es}$ and electron
temperature $T_{\rm e}$ along radial direction. The contribution of
the accretion flows to the spectra includes two parts, the inner
ADAF/RIAF and the outer disk and corona.

For the inner ADAF, we follow the work of Narayan et al. (1995b).
The density $\rho$ and electron Thomson scattering optical depth
$\tau_{\rm es}$ are,
\begin{equation}\label{e:rho_tau}
\begin{array}{l}
\rho=3.79\times10^{-5}\alpha^{-1}c_{1}^{-1}c_{3}^{-1/2}m^{-1}\dot m r^{-3/2}\ \rm g\,cm^{-3},\\
\tau_{es}=12.4\alpha^{-1}c_{1}^{-1}{\dot m}r^{-1/2},
\end{array}
\end{equation}
where
\begin{equation}\label{e:rho_tau}
\begin{array}{l}
{c_1}={(5+2\varepsilon^{'}) \over {3\alpha^2}}g(\alpha,\varepsilon^{'}),\\
\\
{c_3}={2\varepsilon(5+2\varepsilon^{'})\over {9\alpha^2} } g(\alpha,\varepsilon^{'}),\\
\\
{\varepsilon{'}}={\varepsilon\over f}={1\over f} \biggl({{5/3-\gamma}\over {\gamma-1}}\biggr),\\
\\
g(\alpha,\varepsilon^{'})=\biggl[ {1+{18\alpha^2\over (5+2\varepsilon^{'})^{2}}\biggr]^{1/2}-1},\\
\\
\gamma={{32-24\beta-3\beta^2}\over {24-21\beta}},
\end{array}
\end{equation}
with $f$ the fraction of viscosity dissipated energy which is
advected. The energy equation of ions, electrons and the state
equation are  (Narayan et al. 1995b),
\begin{equation}\label{e:Temperature}
\begin{array}{l}
q^{+}=fq^{+}+q^{\rm ie}\\
q^{\rm ie}=q^{-}\\
T_{\rm i}+1.08T_{e}=6.66\times10^{12}\beta c_{3}r^{-1},
\end{array}
\end{equation}
 where $q^{+}$ is the viscous dissipation of energy per unit volume, $q^{\rm ie}$ is the exchange
rate of energy between electrons and ions per unit volume, and
$q^{-}$ is the cooling rate of the electrons per unit volume. The
expression of $q^{+}$, $q^{\rm ie}$ and $q^{-}$ can be found in
Narayan et al (1995b), i.e.,
\begin{equation}\label{e:q}
\begin{array}{l}
q^{+}=1.84 \times 10^{21} \varepsilon{'}c_{3}^{1/2}m^{-2} \dot m r^{-4} \ \rm ergs \ cm^{-3}\ s^{-1},  \\
q^{\rm ie}=5.61 \times 10^{-32} { {n_{e}n_{i}(T_{i}-T_{e})} \over  {K_{2}(1/\theta_{e})} {K_{2}(1/\theta_{i})}}\\
\qquad \times \biggl[ {{2(\theta_{e}+\theta_{i})^{2}+1} \over {(\theta_{e}+\theta_{i})}}K_{1}
\biggl({{\theta_{e}+\theta_{i}} \over {\theta_{e}\theta_{i}}}\biggl)+
2K_{0}\biggl({{\theta_{e}+\theta_{i}} \over {\theta_{e}\theta_{i}}}\biggl)\biggl] \\
\qquad \qquad \qquad \qquad \qquad \qquad \qquad \, \rm ergs \ cm^{-3}\ s^{-1},\\
q^{-}=q_{\rm bresm}^{-}+q_{\rm synch}^{-}+q_{\rm bresm,c}^{-}+q_{\rm synch,c}^{-}+q_{\rm disk,c}^{-}.
\end{array}
\end{equation}
where the $K's$ are modified Bessel functions,
$q_{\rm brsm}^{-}$, $q_{\rm synch}^{-}$, $q_{\rm brsm,c}^{-}$, $q_{\rm synch,c}^{-}$, $q_{\rm disk,c}^{-}$
are the bremsstrahlung cooling rate, synchrotron radiation rate, the cooling rate by the
Comptonization of bremsstrahlung radiation,
synchrotron radiation and the soft photons emitted by the outer disk.
Given $\alpha$, $\beta$, $m$, $\dot m$ and $r$, we can calculate the coronal quantities $T_{\rm e}$, $T_{\rm i}$, $\rho$, $\tau_{\rm es}$ from
equation (15) to equation (18).  Here, for the electron radiation cooling term $q^{-}$, we add the
soft photons which come from the outer truncated thin disk to be scattered by electrons in the
inner ADAF. This cooling term  is
determined by the size of the truncation radius of the thin disk, which can be obtained from our calculation as
described in section 3.1.
With $\rho$, $\tau_{\rm es}$ and $T_{\rm e}$ determined, we can
calculate the emergent spectra from the inner ADAF.

As the first step, we calculate the Bremsstrahlung and synchrotron radiation from the inner ADAF.
Adopting the Eddington approximation and the two-stream approximation (Rybicki \& Lightman 1979; Manmoto et al. 1997), the
radiative diffusion equation in the vertical direction is solved, and the radiative
flux  is given by
\begin{equation}
F_{\rm \nu}={{2\pi}\over \sqrt 3}B_{\rm \nu}[1-\rm exp(-2 \sqrt 3
\tau_{\rm \nu}^{*})],
\end{equation}
where $B_{\rm \nu}$ is the plank function, $\tau_{\nu}^{*}=(\pi^ \frac
{1} {2}) \kappa_{\rm \nu} H$ is the vertical absorption optical depth.
Assuming local thermodynamic equilibrium (LTE), we can write $\kappa_{\rm \nu}= {\chi_{\rm \nu} } / {4\pi B_{\rm \nu}} $,
 $\chi_{\rm \nu} = \chi_{\rm \nu,bresm} + \chi_{\rm \nu,synch}$ is the total emissivity,
$\chi_{\rm \nu,bresm}$ the bremsstrahlung emissivity and $\chi_{\rm
\nu,synch}$  the synchrotron emissivity (for the detailed expression
for $\chi_{\rm \nu,bresm}$ and $\chi_{\rm \nu,synch}$ see Narayan et
al. 1995b). Photons emitted in the Bremsstrahlung and synchrotron
processes can be scattered by electrons in the ADAF (self-Compton
scattering).  With known spectra of soft photons (equation (19)) and
ADAF structure ($T_e$ and $\tau_{\rm es}$), we calculate the local
Compton scattering spectra in terms of multi-scattering method
(Coppi $\&$ Blandford 1990). We integrate this local Compton
radiative spectra from the ISCO to the truncation radius $R_{\rm
tr}$ and get the whole emergent spectra of the inner ADAF. The
method we used here separates the inverse Compton scattering process
from other emission and absorption processes. Because the optical
depth is relative small ($\tau < 1$), this treatment is a good
approximation.

For the outer disk corona, we get the density $\rho$,
Thomson scattering optical depth $\tau_{\rm es}$ and electron temperature $T_{\rm e}$
of the corona from the self-consistent calculation of the disk-corona as described in section 2.
For the disk corona in the vertical direction, there is a thin transition layer
between the cool thin disk and the hot corona. The temperature of the transition layer is about $10^{6.5}$ K,
and the optical depth is relative small, so we neglect the Compton radiation from this layer.
Because of the existence of the underlying thin disk, we add the soft photons flux $F_{\rm \nu,disk}$ from
the thin disk as the initial seed photon flux for the inverse Compton scattering.
The total seed photon flux for the inverse Compton scattering is,
\begin{equation}
\centering
F_{\rm \nu}={{2\pi}\over \sqrt 3}B_{\rm \nu}[1-{\rm exp}(-2 \sqrt 3
\tau_{\rm \nu}^{*})]+F_{\rm \nu, disk},
\end{equation}
where $ F_{\rm \nu, disk}= \pi I_{\nu}=\pi B_{\nu}(T_{\rm eff}),
T_{\rm eff}=\biggl\lbrace\frac{3GM\dot M_{\rm d}}{8\pi R^3 \sigma}
\biggl[1-\sqrt{R_{\rm tr}\over R} \biggr]\biggr\rbrace^{1/4}$,
Using the same method we have described for the ADAF to treat the Comptonization process,
we get the local spectra of the disk and corona at given distance, then we start the integration from
truncation radius $R_{\rm tr}$ to the outer boundary of the accretion disk ($\sim 3000 \,R_{\rm S}$) to
obtain the whole emergent spectra of the outer disk and corona.

Combining the contribution of both the inner ADAF and the outer disk
and corona, we get the total emergent spectra of our model. The
spectra are plotted in Figure 4 for mass accretion rates $\dot M=$
0.001, 0.003, 0.005, 0.008, $0.01 \,\dot M_{\rm Edd} $. For a lower
mass accretion rate $\dot M=$ 0.001 $ \dot M_{\rm Edd}$, the disk
truncates at 471 $R_{\rm S}$. Because of the large truncation
radius, the radiation from the inner ADAF dominates over that of the
outer disk and corona, as shown by the dotted curve in the first
panel of Figure 5. The first peak of this spectrum is dominated by
the self-absorbed synchrotron radiation of the ADAF. The hard X-ray
spectra is dominated by the thermal bremsstrahlung of the ADAF. For
$\dot M=$ 0.003 $\dot M_{\rm Edd}$, the spectrum is plotted in the
upper, right panel of Figure 5. The thin disk moves inwards and
truncates at around $200 R_{\rm S}$. In this case, the radiation
from the inner ADAF still dominates over that of the outer disk and
corona. However, the contribution of the truncated disk and corona
to the first peak of the spectrum increases, and the Compton
radiation of the ADAF becomes more important for the hard X-ray
spectrum. With increase of the mass accretion rate, say,  for $\dot
M=$ 0.008, 0.01 $\dot M_{\rm Edd}$, the thin disk moves inwards
further, and truncates at 78.5 $R_{\rm S}$, 40 $R_{\rm S}$
respectively. The first peak of the spectra is dominated by the
radiation from the truncated disk, and the second peak around 100
keV in X-ray band is dominated by the Compton radiation of the inner
ADAF, as shown by the lower panels of Figure 5.
Here, we note that
although the truncation radius is small (40 $R_{\rm S}$) for
$\dot{M} = 0.01 \dot{M}_{\rm Edd}$, the emission of ADAF is still
more luminous than that of lower $\dot{M}$ with large truncation
radius. This can be understood as, firstly, the luminosity of ADAF
is proportional to $\dot M^2$, so with increase of the accretion
rate the luminosity increases; secondly, most potential energy of
ADAF is released in the most inner region around the black hole
within the framework of ADAF self-similar solution.

The variation of the hard X-ray spectra with the mass accretion rates is plotted in Figure 6.
We demonstrate that the hard X-ray photon index $\Gamma_{\rm 3-25 \,keV}$ is anti-correlated with
the mass accretion rate. This is because, with increase of the accretion rate, the Comptonization of the
synchrotron and bremsstrahlung photons by the hot electrons becomes dominant. Therefore, the Thomson
optical depth and hence Compton $y$-parameter increases, resulting in a harder X-ray spectra.

If $\dot M > 0.01 \,\dot M_{\rm Edd}$,  the thin disk extends down to the ISCO,
and above the thin disk only a weak corona exits as a consequence of strong external Compton scattering.
The emission property in this case will be studied in the future.

\section{Comparison with XTE 1118+480}
Observation shows there is a strong anti-correlation between the Eddington ratio
$\xi$ and the hard X-ray photon index
$\Gamma$ for the black hole X-ray binary XTE 1118+480 at low Eddington ratio $\xi < 0.01$.
We compare this correlation with the prediction of disk and corona model.

McClintock et al. (2001) and Wagner et al. (2001) independently confirmed the orbital period to be
4.1 hr (as previously suggested by Cook et al. 2000; Patterson 2000; Uemura et al. 2000;
Dubus et al. 2001) and determined the value of the mass function to be $f(M)$ $\simeq$
$6.0 M_\odot$ (Esin et al. 2001). The distance was estimated to
be 1.8 $\pm$ 0.6 kpc (McClintock et al. 2001) and was took as $d=2.9$ kpc to model the multiwavelength spectrum (Esin et al. 2001).
By taking $M=6M_{\odot}$, viscosity parameter $\alpha=0.3$,
the ratio of gas pressure to total pressure $\beta=0.5$, we show our theoretical relation between the
Eddington ratio $\xi \equiv L_{\rm 0.5-25 \,keV}/L_{\rm Edd}$
and the hard X-ray photon index $\Gamma_{3-25 \,\rm keV}$ in Figure 7 for a series of accretion rate
$\dot M=0.001,0.003,0.005,0.008$, $0.01 \,\dot M_{\rm Edd}$.  The pluses are the observational data (Wu et al. 2008) for taking
the black hole mass as $M=6M_{\odot}$, and the distance $d =2.9 $ kpc (Esin et al. 2001)
(Note: We get the observational data from Wu et al. (2008). In the work of Wu et al. (2008) the black hole
mass $M=6.1M_{\odot}$ and the distance $d$=1.8 kpc are adopted to obtain $\xi$. We recalculate the value
of $\xi$ by taking $M=6M_{\odot}$ and the distance $d$ = 2.9 kpc).
Because $\xi$ is related to the distance, we take different estimated distance for comparisons.
The diamonds are for $d$ = 1.8+0.6 kpc (McClintock et al. 2001),
and the pluses  for $d$=1.8 kpc (Wu et al. 2008).
One can see our theoretical results for the bigger distance $d$=2.9 kpc can fit the observational data
better than that of $d$=1.8 kpc. Due to the uncertainty of the measurement to the distance,
our theoretical result can only reproduce this tendency of anti-correlation to some extent.

\section{Discussion}
\subsection{The Effect of the Viscosity Parameter and the Ratio of
Magnetic Pressure to Gas Pressure}
Throughout our calculations, the viscosity parameter  and the ratio of
magnetic pressure to gas pressure are fixed to $\alpha=0.3$ and $\beta=0.5$.
Varying these parameters leads to a change of truncation radius at any given accretion rate
(Qian et al. 2007); Qiao \& Liu 2009). Nevertheless, the spectrum is only slightly affected by the truncation
 radius since the total emitted spectrum is dominated by the inner ADAF, in particularly in the case of large
 truncation radii. Only when the truncation radius is as small as $\sim 10R_S$, where the Compton cooling of
 the soft photons from the outer disk to the ADAF becomes the dominated cooling mechanism,
 can the spectral index be largely affected. However, such a small truncation radius is very difficult to reach according to our model.

The viscosity parameter can also affect the critical transition rate (Meyer-Hofmeister et al. 2001; Qiao \& Liu 2009).
For a larger value, $\alpha>0.3$,  the hard state extends to higher accretion rates, $\dot m>0.01$. In this case, the
distribution of photon index as a function of accretion rate, as shown in Figure 6,  cannot be changed much at low
accretion rates ($\dot m< 0.01$)   or low luminosities since the truncation radius is large and hence the
emitting spectrum is dominated by the inner ADAF. However, at high accretion rates (from $\dot m=0.01$ up to
the state transition rate), the disk is truncated at smaller radii, which could result in softer photon index
than that shown in Figure 6. For a smaller viscosity,  $\alpha<0.3$, the disk is truncated at larger distances.
The spectrum is hardly affected at any given accretion rates. Only the lower limit of the photon index (see figure 6)
is larger because the predicted transition rate is lower.

\subsection{The Challenge to the Disk Truncation}
The observation of  XTE J1650-500 (Miller et al. 2002; Miniutti et
al. 2004) and GX 339-4 (Miller et al. 2006a; b) demonstrates that
the smeared iron line and reflection feature in the low/hard state
are inconsistent with disk truncation. The claimed broadening and
skewed iron line, which may be caused by the Doppler shift and
strong general relativistic effect, require that the thin disk
extends to the ISCO (Fabian et al. 2000; Miller 2007; Done et al.
2007). Recently, the condensation of matter from a hot corona to a
cool, optically thick innermost disk is proposed for black hole
transient systems in the low/hard state (Meyer et al. 2007; Liu et
al. 2007; Taam et al. 2008). This model may alleviate the
contradiction that a thin disk exists near ISCO in the low/hard
state.
 Present observational technique can not directly diagnose  whether the thin disk truncates at some radius or not
 (Young et al. 2000; $\rm Gierli\acute{n}ski$ et al. 1997).
Further more advanced instruments are expected to solve this problem.
\section{Conclusion}
In this work, we investigate the spectral features of accretion flows composed of an outer cool,
optically thick disk and inner hot, optically thin, advection dominated accretion flows (ADAF) within the
framework of disk and corona with mass evaporation.
The disk is truncated at the distance where  the evaporation rate equals to the accretion
rate ($\dot m_{\rm evap}(r_{\rm tr})=\dot m$).
We show the disk evaporation mechanism can connect the outer disk and the inner ADAF/RIAF smoothly.
We calculate the emergent spectra of the disk and corona with mass evaporation in the low/hard state for different
mass accretion rates. From the theoretical spectra, we find an anti-correlation
between the Eddington ratio $\xi =L_{\rm 0.5-25 \,keV}/L_{\rm Edd}$ and the hard X-ray photon index
$\Gamma_{\rm 3-25 \,keV}$, which agrees well with the observations for the black hole X-ray binary XTE J1118+480.
\\

We would like to thank Marat Gilfanov, F. Meyer and E. Meyer-Hofmeister for their promising discussion.
Erlin Qiao thanks the support of Doctoral Promotion Program between Chinese Academy
of Sciences and Max Planck Society. In addition, this work is supported by the National Nature Science
Foundation of China (grants 10533050 and 10773028) and by the National Basic Research Program of
China-973 Program 2009CB824800.

\begin{figure*}
\includegraphics[width=85mm,height=70mm,angle=0.0]{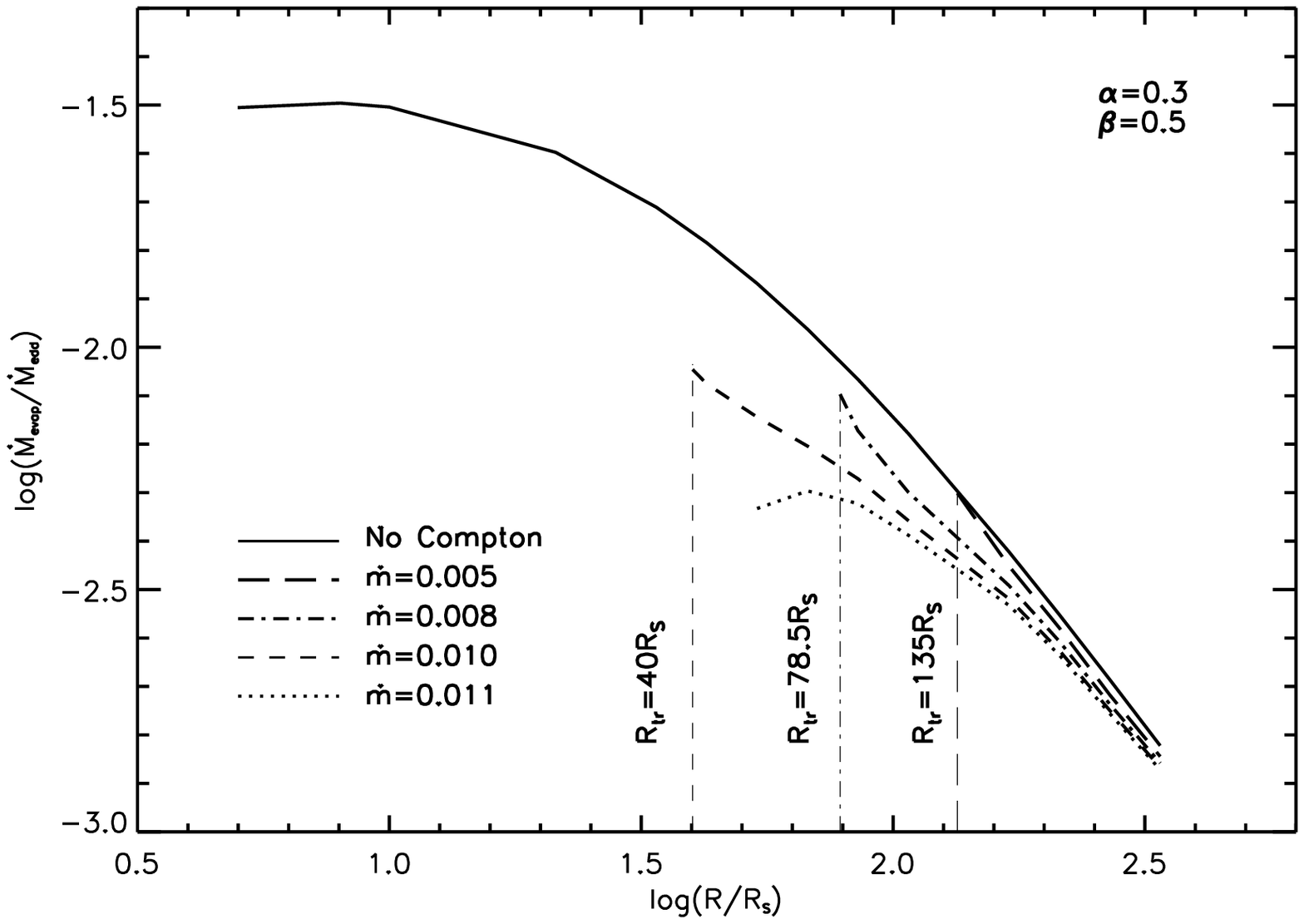}
\caption{The evaporation rates along radial direction for different mass accretion rates.
The solid curve corresponds to the evaporation rate without Compton
cooling of the soft photons emitted by the thin disk.
The long-dashed line denotes the evaporation rate distribution
along the radial direction for $\dot M=$ 0.005 $\dot M_{\rm Edd}$,
the dot-dashed line for $\dot M=$ 0.008 $\dot M_{\rm Edd}$, the short-dashed line
for $\dot M=$ 0.01 $\dot M_{\rm Edd}$, and the dotted line for $\dot M=$ 0.011 $\dot M_{\rm Edd}$.
For $\dot M=$0.005 $\dot M_{\rm Edd}$, $\dot M=$ 0.008 $\dot M_{\rm Edd}$ and $\dot M=$ 0.01 $\dot M_{\rm Edd}$
the thin disk truncates at around 135 $R_{\rm S}$, 78.5 $R_{\rm S}$ and 40 $R_{\rm S}$ respectively.
For $\dot M= 0.011 \dot M_{\rm Edd}$, the thin disk extends to the ISCO.
In our calculation, $M=6M_\odot$, $\alpha=0.3$ and $\beta=0.5$ are adopted.}
\end{figure*}

\begin{figure*}
\includegraphics[width=85mm,height=70mm,angle=0.0]{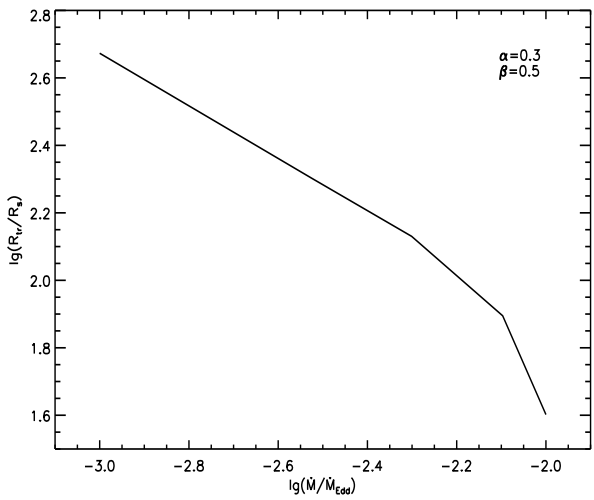}
\caption{The dependence of the truncation radius on the mass
accretion rate. $M=6M_\odot$, $\alpha=0.3$ and $\beta=0.5$ are adopted.}
\end{figure*}

\begin{figure*}
\includegraphics[width=85mm,height=70mm,angle=0.0]{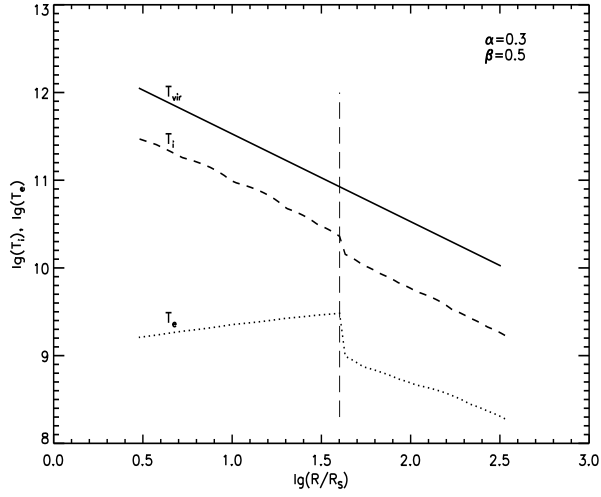}
\caption{The temperature versus the distance from the black hole for $\dot M =0.01 \dot M_{\rm Edd}$.
The solid line is for the virial temperature. The dashed line and the dotted line are the ion temperature and the
electron temperature distribution respectively.
The lines in the left part of the vertical long-dashed line is for the inner
self-similar solution of ADAF, and the right part is for the
outer corona solution.
$M=6M_\odot$, $\alpha=0.3$ and $\beta=0.5$ are adopted.}
\end{figure*}

\begin{figure*}
\includegraphics[width=85mm,height=70mm,angle=0.0]{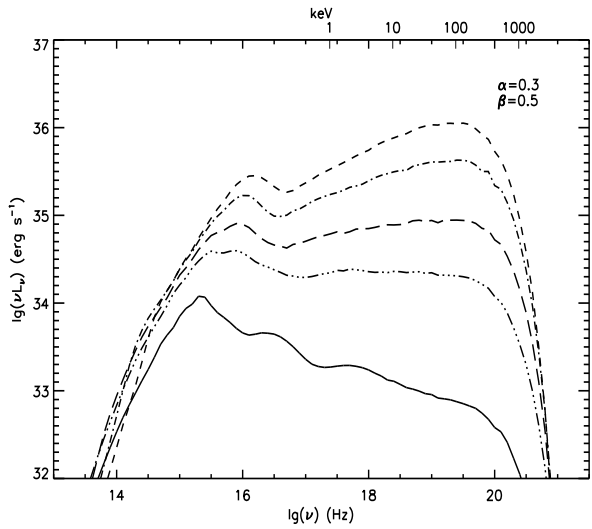}
\caption{The emergent spectra of disk and corona with mass evaporation.
The curves from the bottom up are the emergent spectra for $\dot M=$ 0.001,
0.003, 0.005, 0.008, 0.01 $\dot M_{\rm Edd} $ respectively.
$M=6M_\odot$, $\alpha=0.3$ and $\beta=0.5$ are adopted.}
\end{figure*}

\begin{figure*}
\includegraphics[width=85mm,height=70mm,angle=0.0]{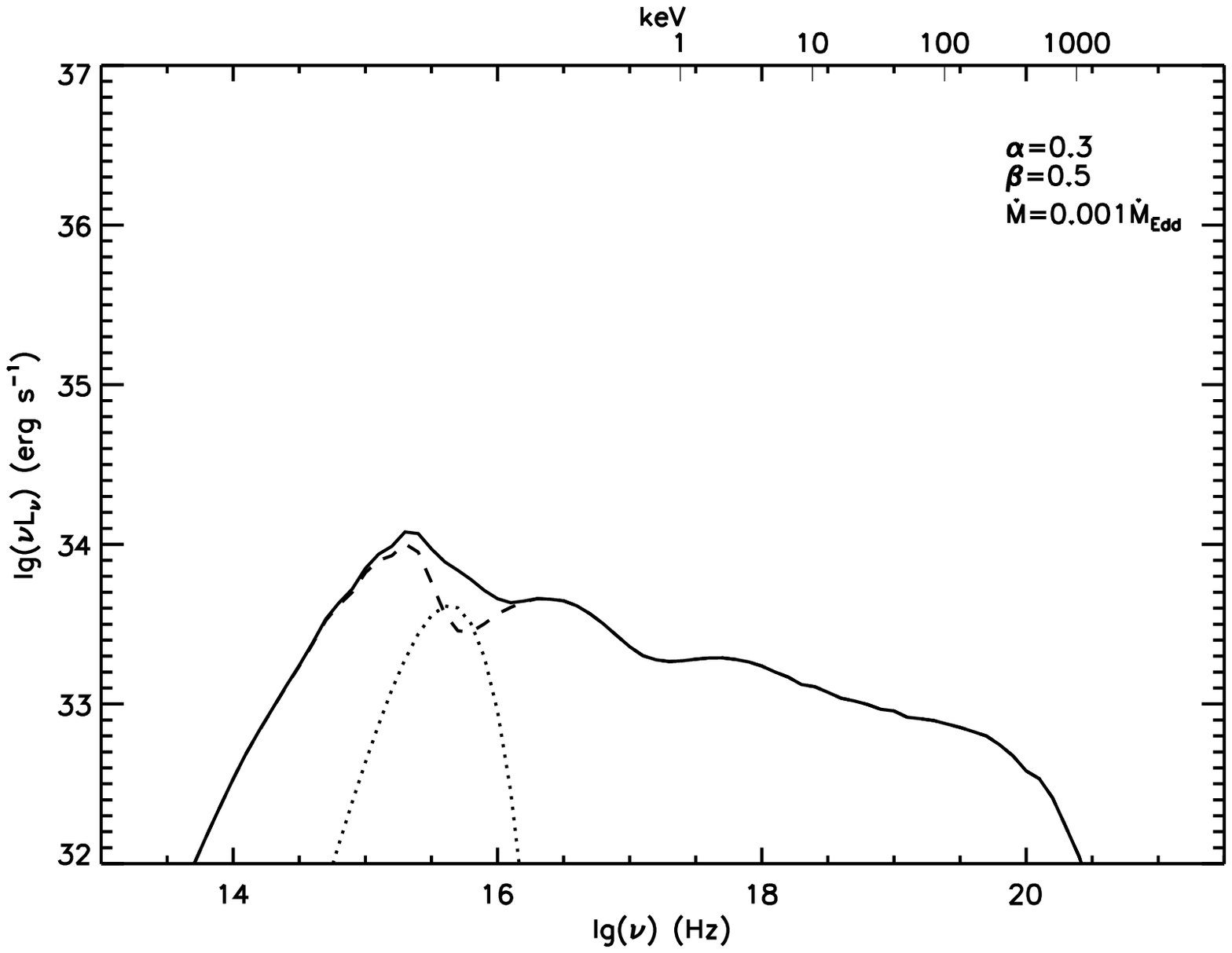}
\includegraphics[width=85mm,height=70mm,angle=0.0]{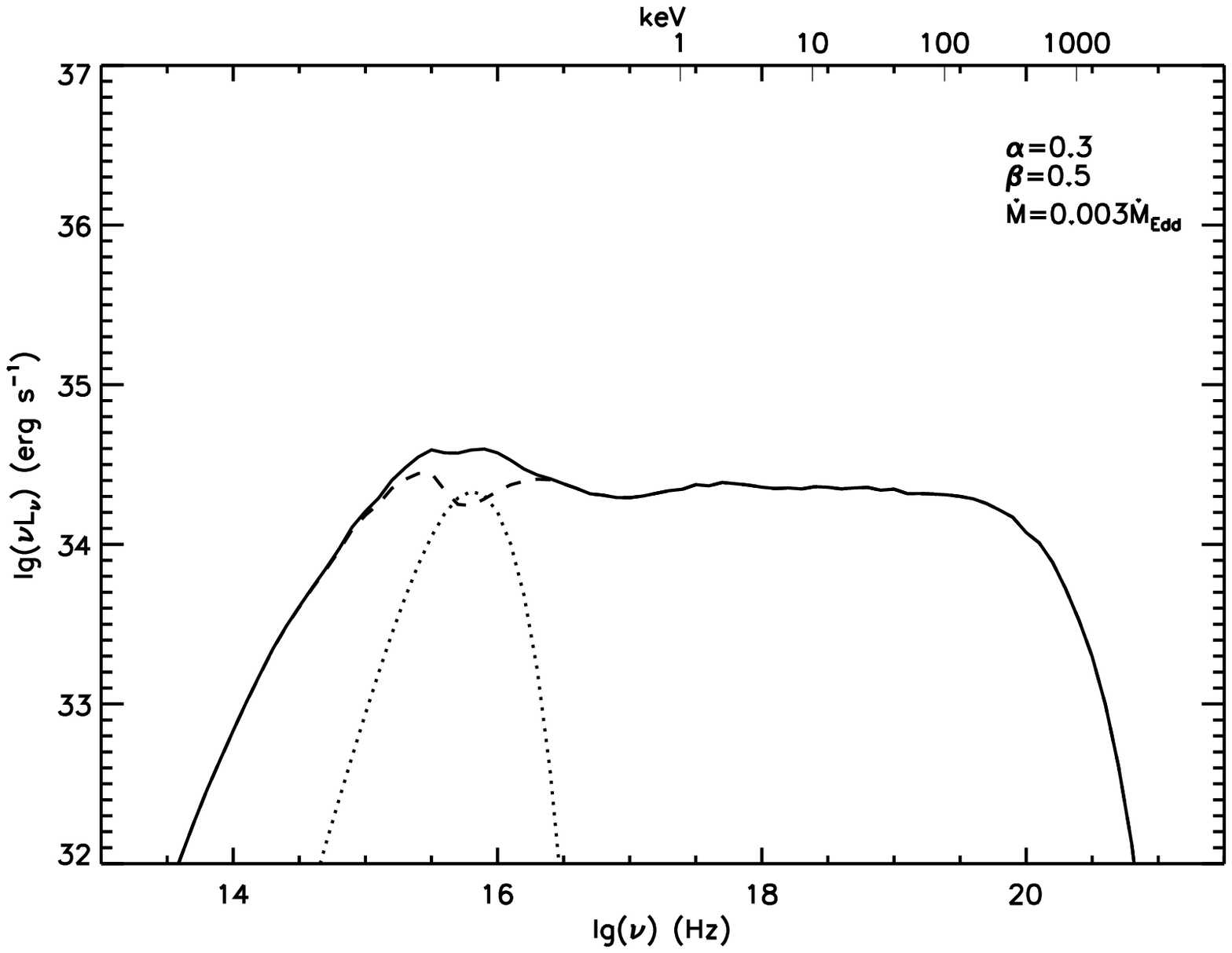}
\includegraphics[width=85mm,height=70mm,angle=0.0]{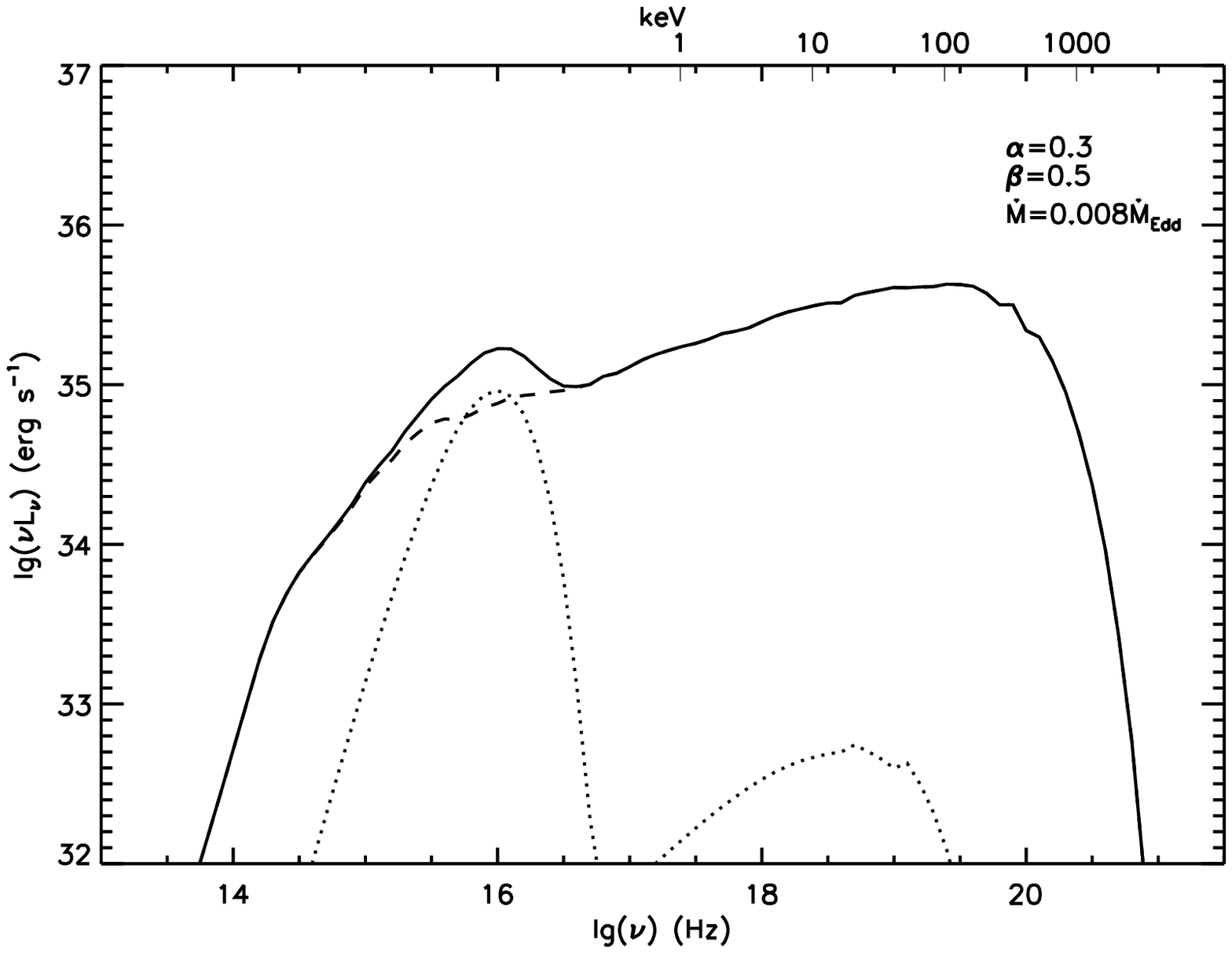}
\includegraphics[width=85mm,height=70mm,angle=0.0]{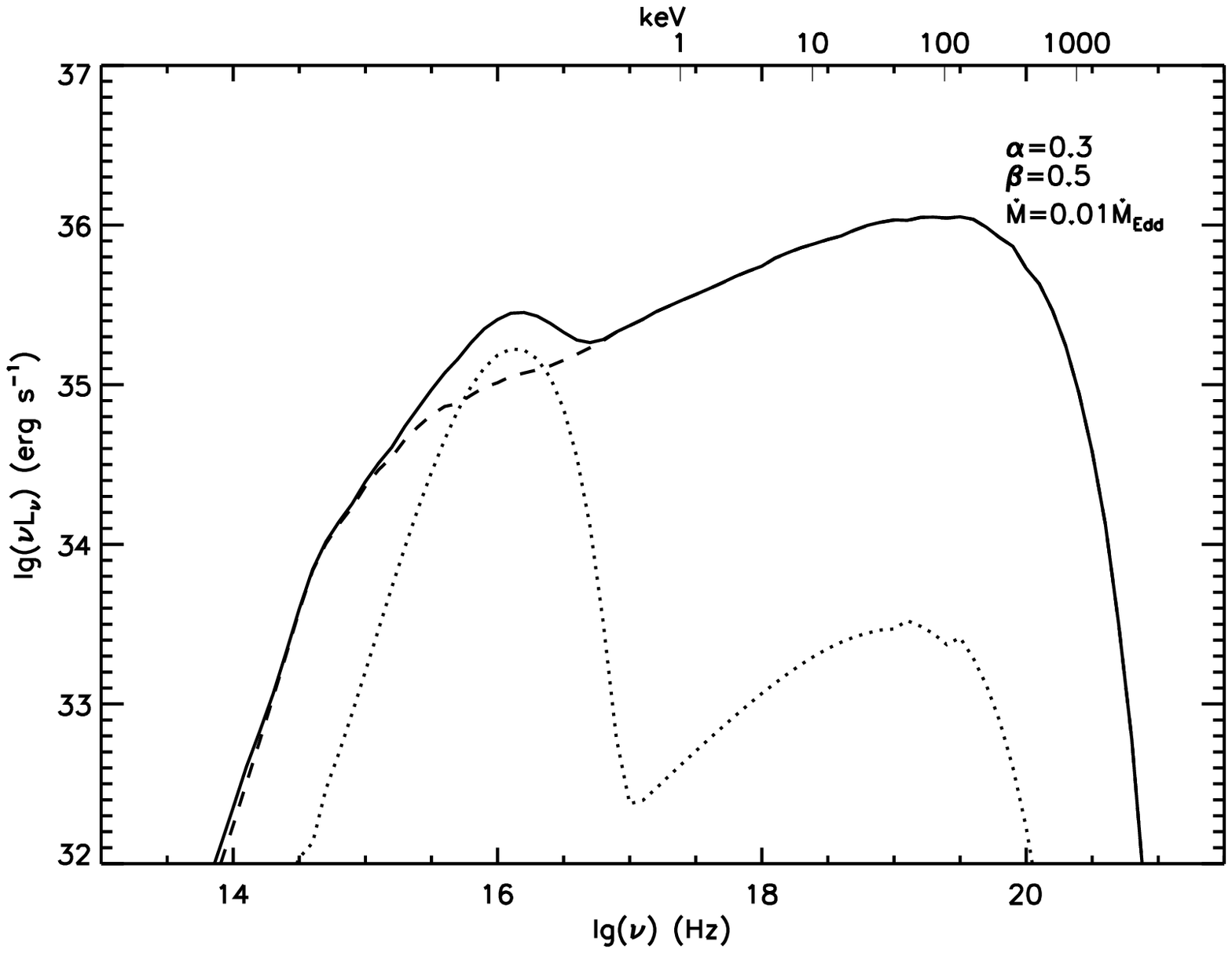}
\caption{The emergent spectra of disk and corona with mass evaporation for different mass accretion rates
$\dot M = 0.001, 0.003, 0.008, 0.01$ $\dot M_{\rm Edd}$.
In each plot, the dashed line is the spectra of the inner ADAF and the dotted line
is the spectra of the outer disk-corona. The solid line is the sum. $M=6M_\odot$, $\alpha=0.3$ and
$\beta=0.5$ are adopted.}
\end{figure*}

\begin{figure*}
\includegraphics[width=85mm,height=70mm,angle=0.0]{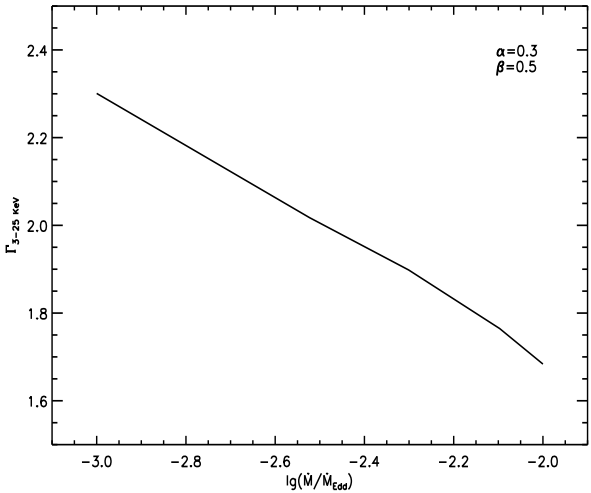}
\caption{The dependence of the hard X-ray photon index on the mass accretion rate.
$M=6M_\odot$, $\alpha=0.3$ and $\beta=0.5$ are adopted.}
\end{figure*}

\begin{figure*}
\includegraphics[width=85mm,height=70mm,angle=0.0]{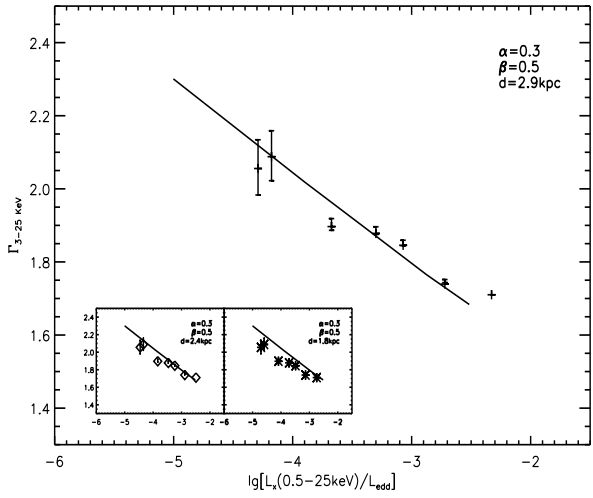}
\caption{The anti-correlation between the Eddington ratio $\xi$ and the hard X-ray photon index
$\Gamma_{\rm 3-25 \,keV}$. The solid line are our theoretical prediction
assuming $M=6M_\odot$, $\alpha=0.3$, $\beta=0.5$.
The pluses denote the observational data of X-ray source XTE J1118+480
for distance $d$ = 2.9 kpc. The diamonds in the first inset are for $d$ = 1.8+0.6 kpc, and the
asterisks in the second inset are for $d$ = 1.8 kpc.}
\end{figure*}

\end{document}